\documentclass[prd,preprint,superscriptaddress,amsmath,amssymb,nofootinbib]{revtex4}
\usepackage{graphicx}
\usepackage{dcolumn}
\usepackage{bm}
\usepackage{amssymb}
\usepackage{amsmath}
\usepackage{epsfig}    
\usepackage{color}
\usepackage{slashed}
\usepackage{hhline}

\def\be{\begin{equation}}
\def\ee{\end{equation}}
\newcommand{\bea}{\begin{eqnarray}}
\newcommand{\eea}{\end{eqnarray}}
\newcommand{\nn}{\nonumber}

\begin{document}

\begin{flushright}{KIAS-P17075} \end{flushright}

\title{Left-handed and right-handed $U(1)$ gauge symmetry}

\author{Takaaki Nomura}
\email{nomura@kias.re.kr}
\affiliation{School of Physics, KIAS, Seoul 02455, Korea}

\author{Hiroshi Okada}
\email{macokada3hiroshi@cts.nthu.edu.tw}
\affiliation{Physics Division, National Center for Theoretical Sciences, Hsinchu, Taiwan 300}

\date{\today}

\begin{abstract}
 We propose a model with the left-handed and right-handed continuous Abelian gauge symmetry; $U(1)_L\times U(1)_R$. Then three right-handed neutrinos are naturally required to achieve $U(1)_R$ anomaly cancellations, while several mirror fermions are also needed to do  $U(1)_L$ anomaly cancellations.  Then we formulate the model, and discuss its testability of the new gauge interactions at collider physics such as  the large hadron collider (LHC) and the international linear collider (ILC).  In particular, we can investigate chiral structure of the interactions by the analysis of forward-backward asymmetry based on polarized beam at the ILC.
\end{abstract}
\maketitle

\section{Introduction}
A new right-handed gauge symmetry $U(1)_R$~\cite{Geng:1993ja, Nomura:2016emz, Nomura:2016pgg, Nomura:2017ezy, Nomura:2017tih, Chao:2017rwv} is one of the 
promising candidates naturally to accommodate the three right-handed neutrinos to achieve the anomaly cancellations, whose nature is the same as a theory of $B-L$ gauge symmetry~\cite{Mohapatra:1980qe}.
In addition, it is a verifiable and characteristic candidate to be tested by phenomena at current and future collider experiments such as international linear collider (ILC)~\cite{ilc}
by measuring several differential cross sections for purely polarized initial states as well as the large hadron collider (LHC).
In particular, the new gauge interaction of $U(1)_R$ can be distinguished from that of $U(1)_{B-L}$ models~\cite{Nomura:2017tih}.
This arises from the chiral asymmetry between right-handed and left-handed fermions in the new gauge interactions of a theory.
However one might think why only the $U(1)_R$ gauge symmetry is there, and/or what about the left-handed type gauge symmetry $U(1)_L$ under which only left-handed fermions are charged.
Actually the minimal $U(1)_R$ model requires its rather large breaking scale $\sim{\cal O}(20)$ TeV, due to a few number of parameters in the gauge sector~\cite{Nomura:2017tih, Chao:2017rwv}, which can be relaxed considering $U(1)_L$ gauge symmetry.
It is thus interesting to consider $U(1)_L$ gauge symmetry together with the $U(1)_R$ symmetry and discuss how to distinguish the two types of gauge interactions at the collider experiments.  

In this paper, we extend the minimal $U(1)_R$ gauge symmetry into the $U(1)_L\times U(1)_R$ and construct a consistent model in a minimal way.
Then we need exotic quarks and leptons in order to cancel the new gauge anomalies, two Higgs doublet fields to induce nonzero SM fermion masses, and two SM singlet scalar fields with new $U(1)$ charges to break the new gauge symmetries and to provide masses of exotic fermions.
As a result, breaking scale of $U(1)_L\times U(1)_R$ can be lower than the case with only $U(1)_R$ due to additional parameters and degrees of freedom in the gauge boson sector.
Then we formulate each of fermion sector, Higgs boson sector, vector gauged boson sector, as well as neutrino sector.
We show that  the Yukawa interaction among two Higgs doublets and SM fermions is that of the type-II two Higgs doublet model (THDM) due to the new gauge symmetry. 
In addition we discuss the possibility of testing the new gauge interactions at current and future collider such as LHC and ILC and of distinguishing differences between our model and the others.

 This letter is organized as follows.
In Sec. II, we introduce our model, and formulate Higgs sector, neutral gauge sector, neutrino sector, and interacting terms.
In Sec. III,  we discuss phenomenologies of new neutral gauge boson at colliders.
 Finally we devote the summary of our results and the conclusion in Sec. IV.

\section{Model setup and Constraints}
\begin{table}[t!]
\begin{tabular}{|c||c|c|c|c|c|c|c|c|c|c|}\hline\hline  
& ~$Q_L^a$~& ~$u_R^a$~  & ~$d_R^a$~& ~$L_L^a$~& ~$e_R^a$~& ~$\nu_R^a$~& ~$Q'^a_L$~& ~$Q'^a_R$~& ~$L'^a_L$~& ~$L'^a_R$ \\\hline\hline 
$SU(3)_C$ & $\bm{3}$  & $\bm{3}$ & $\bm{3}$ & $\bm{1}$ & $\bm{1}$ & $\bm{1}$ & $\bm{3}$ & $\bm{3}$& $\bm{1}$ & $\bm{1}$  \\\hline 
$SU(2)_L$ & $\bm{2}$  & $\bm{1}$  & $\bm{1}$  & $\bm{2}$  & $\bm{1}$  & $\bm{1}$  & $\bm{2}$ & $\bm{2}$  & $\bm{2}$ & $\bm{2}$   \\\hline 
$U(1)_Y$   & $\frac16$ & $\frac23$ & $-\frac13$ & $-\frac12$  & $-1$ & $0$  & $\frac16$  & $\frac16$ & $-\frac12$  & $-\frac12$\\\hline
$U(1)_{L}$   & $\ell$ & $0$ & $0$ & $\ell$  & $0$  & $0$  & $0$  & $\ell$  & $0$  & $\ell$\\\hline
$U(1)_{R}$   & $0$ & $r$ & $-r$   & $0$  & $-r$  & $r$  & $0$  & $0$  & $0$  & $0$\\\hline
$Z_2$   & $+$ & $+$ & $+$   & $+$  & $+$  & $+$  & $-$  & $-$  & $-$  & $-$\\\hline
\end{tabular}
\caption{ 
Charge assignments of the our fields
under $SU(3)_C\times SU(2)_L\times U(1)_Y\times U(1)_{L} \times U(1)_{R}$ with $r,\ell\neq0$, where their upper indices $a$ are the number of family that run over $1-3$.}
\label{tab:1}
\end{table}

\begin{table}[t!]
\centering {\fontsize{10}{12}
\begin{tabular}{|c||c|c|c|c|}\hline\hline
  Bosons  &~ $H_u$  ~ &~ $H_d$~ &~ $\varphi_L$ &~ $\varphi_R$ \\\hline
$SU(3)_C$ & $\bm{1}$   & $\bm{1}$  & $\bm{1}$ & $\bm{1}$ \\\hline 
$SU(2)_L$ & $\bm{2}$  & $\bm{2}$ & $\bm{1}$ & $\bm{1}$  \\\hline 
$U(1)_Y$ & $\frac12$ & $\frac12$ & $0$ & $0$    \\\hline
 $U(1)_{L}$ & $-\ell$ & $\ell$ & $\ell$  & $0$  \\\hline
  $U(1)_{R}$ & $r$ & $r$ & $0$  & $2r$  \\\hline
    $Z_2$ & $+$ & $+$ & $+$  & $+$  \\\hline
\end{tabular}%
} 
\caption{Charge assignments in scalar sector. }
\label{tab:2}
\end{table}

In this section we review our model based on $U(1)_L \times U(1)_R$ symmetry where the left- and right-handed SM fermions are charged under $U(1)_L$ and $U(1)_R$ respectively.
In the fermion sector,
we add three families of right-handed Majorana fermions $\nu_R^a$($a=1-3$) to cancel the $U(1)_R$ anomalies among SM fermions, which are the same assignments as ref.~\cite{Nomura:2017tih, Chao:2017rwv} due to their zero charges for the other exotic fermions; $Q'$ and $L'$. Three $Q'$ and $L'$ are also introduced to cancel the $U(1)_L$ anomalies among SM fermions, whose assignments are similar to the one of ref.~\cite{Nomura:2017tzj}. The other nontrivial anomalies between $U(1)_L$ and $U(1)_R$ arise from $[U(1)_R]^2 U(1)_L$, $U(1)_R [U(1)_L]^2$, and $U(1)_Y U(1)_R U(1)_L$,  but these are automatically anomaly free because all the fermions have zero charge under either $U(1)_L$ or $U(1)_R$. All the fermion contents and their assignments are summarized in Table~\ref{tab:1}.
In the scalar sector, we have to extend Higgs sector as THDM in order to provide the SM fermion masses for up- and down-type quark sector, which is a direct consequence of the extension to $U(1)_L$ gauge symmetry. 
In addition, we introduce two isospin singlet scalar fields $\varphi_L$ and $\varphi_R$ to induce the spontaneously symmetry breaking 
of $U(1)_L$ and $U(1)_R$, respectively. These singlet scalar fields also play a role in providing the masses for $Q'$ and $L'$.
All the scalar contents and their assignments are summarized in Table~\ref{tab:2}.
Note that $Z_2$ symmetry is assigned in order to evade mixing mass terms between the SM fermions and exotic fermions such as  $\bar Q_L Q'_R$ and $\bar L_L L'_R$ just for simplicity.~\footnote{In this sense, $Z_2$ is not so important, and one can remove this symmetry without conflict of crucial experimental constraints.}
The relevant Yukawa interactions under these symmetries are given by 
\begin{align}
-{\cal L}_{Y}
&=  (y_u)_{ab} \bar Q^a_L \tilde H_u u^b_R +  (y_d)_{ab} \bar Q^a_L H_d d^b_R+ (y_\ell)_{ab} \bar L^a_L H_d e^b_R
+ (y_D)_{ab} \bar L^a_L\tilde H_u \nu^b_R
\nn\\&
 +  (y_\nu)_{aa} \bar\nu^{aC}_R \nu^a_R \varphi_R^* +(y_Q')_{aa}\bar Q'^{a}_L Q'^{a}_R \varphi_L^*
 +(y_L')_{aa}\bar L'^{a}_L L'^{a}_R \varphi_L^*
+ {\rm h.c.}, \label{Eq:yuk} 
\end{align}
where $\tilde H\equiv i\sigma_2H$, and upper indices $(a,b)=1$-$3$ are the number of families, and $y_\nu$, $y_Q'$, and $y_L'$ can be diagonal matrix without loss of generality due to the phase redefinitions of corresponding fermions. Notice that our Yukawa interactions for the SM fermions are the same as that in the Type-II THDM.
In addition, the scalar potential in our model is written as 
\begin{align}
{\cal V} & = m_1^2 |H_u|^2 + m_2^2 |H_d|^2 + m_{\varphi_L}^2 |\varphi_L|^2 + m_{\varphi_R}^2 |\varphi_R|^2  \nonumber \\
& + \frac{\lambda_1}{2} |H_u|^4 + \frac{\lambda_2}{2} |H_d|^4 + \lambda_3 |H_u|^2 |H_d|^2 + \lambda_4 |H_u^\dagger H_d|^2 \nonumber + \lambda_L |\varphi_L|^4 + \lambda_R |\varphi_R|^4 \\
& + \lambda_{H_u \varphi_L} |H_u|^2 |\varphi_L|^2 +  \lambda_{H_u \varphi_R} |H_u|^2 |\varphi_R|^2 
+  \lambda_{H_d \varphi_L} |H_d|^2 |\varphi_L|^2 +  \lambda_{H_d \varphi_R} |H_d|^2 |\varphi_R|^2 \nonumber \\
& + \lambda_{LR} |\varphi_L|^2 |\varphi_R|^2 + \lambda_0 \left[(H^\dag_u H_d)\varphi^{*2}_L + {\rm c.c.}\right],
\label{Eq:pot}
\end{align} 
where the last term is non-trivial in the potential and we assume all the couplings are real.
Here we note that $H^\dagger_u H_d$ and $(H^\dagger_u H_d)^2$ terms are absent in the THD sector due to the exotic $U(1)_L$ gauge symmetry.

\subsection{Scalar sector}

The scalar fields are parameterized as 
\begin{align}
&H_u =\left[\begin{array}{c}
w^+_u\\
\frac{v_u + r_u +i z_u}{\sqrt2}
\end{array}\right],\ 
H_d =\left[\begin{array}{c}
w^+_d\\
\frac{v_d + r_d +i z_d}{\sqrt2}
\end{array}\right],\ 
\varphi_L=
\frac{v_L+ r_L + iz_L }{\sqrt2},\
\varphi_R=
\frac{v_R+ r_R + iz_R }{\sqrt2},
\label{component}
\end{align}
where the singly charged sector $w^\pm_{u,d}$ can be considered as the same manner in the THDM~\cite{Gunion:1989we}.
In the singly charged boson sector, we have two by two mass matrix squared $M^2_C$, and diagonalized by orthogonal mixing matrix $O_C$ as $(M_C^{\rm diagonal})^2 \equiv O_C M^2_C O_C^T$, therefore $[w^\pm,H^\pm]^T=O_C^T [w_u^\pm,w_d^\pm]^T$,
where $w^\pm$ is absorbed by charged gauge boson $W^\pm$. These analytical forms are also found to be as follows:
\begin{align}
& M^2_C = 
\left[\begin{array}{cc}
-\frac{v_d (\lambda_0 v_L^2+\lambda_4 v_uv_d)}{2 v_u} &  \frac{\lambda_0 v_L^2+\lambda_4 v_uv_d}{2}\\ 
 \frac{\lambda_0 v_L^2+\lambda_4 v_uv_d}{2} & -\frac{v_u (\lambda_0 v_L^2+\lambda_4 v_uv_d)}{2 v_d}  \\ 
\end{array}\right],\\
& O_C =
\left[\begin{array}{cc}
c_\beta & s_\beta  \\ 
- s_\beta &  c_\beta  \\ 
\end{array}\right],\quad \tan \beta \equiv \frac{v_u}{v_d}, \\
& (M_C^{\rm diagonal})^2 
=
{\rm Diag}\left[0,  - \frac{\lambda_0 v_L^2}{2 s_\beta c_\beta} - \frac{v^2}{2} \lambda_4 \right],
\end{align}
where $c_\beta(s_\beta) = \cos \beta (\sin \beta)$ and  $v\equiv \sqrt{v_u^2+  v_d^2}$.

As for the CP-even sector in the basis of $[r_u,r_d,r_L,r_R]^t$,  we have four by four mass matrix squared $M^2_R$, and diagonalized by orthogonal mixing matrix $O_R$ as $D[h_1,h_2,h_3,h_4] \equiv O_R M^2_R O_R^T$, therefore $[r_u,r_d,r_L,r_R]^t=O_R^T [h_1,h_2,h_3,h_4]^t$. Here we identify $h_1\equiv h_{SM}$.

In the similar way for the CP-even sector, we have four by four mass matrix squared $M^2_I$.
Since we do not have $H^\dagger_u H_d$ and $(H^\dagger_u H_d)^2$ terms in the THD sector, the non-zero physical CP-odd mass term comes from
$(H^\dag_u H_d)\varphi^{*2}_L + {\rm c.c.}$, associated with coupling $\lambda_0$.
Then the mass matrix is diagonalized by orthogonal mixing matrix $O_I$ as $D[z_1,z_2,z_3,z_4] \equiv O_I M^2_I O_I^T$, therefore $[z_u,z_d,z_L,z_R]^t=O_R^T [z_1,z_2,z_3,z_4]^t$,
where $z_1,z_2,z_3$ are massless bosons. These are analytically found to be as follows:
\begin{align}
& M^2_I =
\lambda_0 
\left[\begin{array}{ccc}
-\frac{v_d v_L^2}{2 v_u} &  \frac{v_L^2}{2} & -{v_d v_L} \\ 
  \frac{v_L^2}{2} &  -\frac{v_u v_L^2}{2 v_d} & v_u v_L \\ 
-{v_d v_L} &   v_u v_L & -2 v_u v_d \\ 
\end{array}\right],\\
& O_I =
\left[\begin{array}{ccc}
-\frac{2v_u}{v'} & 0 & \frac{v_L}{v'} \\ 
\frac{v_u v_L^2}{\sqrt{v'^2(v_u^2 v_L^2+v_d^2 v'^2)}} & \frac{ v_d  v'}{\sqrt{v_u^2 v_L^2+v_d^2 v'^2)}} & 
\frac{2v_u^2 v_L}{\sqrt{v'^2(v_u^2 v_L^2+v_d^2 v'^2)}}  \\ 
\frac{v_d v_L}{\sqrt{4v_u^2 v_d^2+v_L^2 v^2}} & \frac{v_u v_L}{\sqrt{4v_u^2 v_d^2+v_L^2 v^2}} & 
\frac{2v_u v_d}{\sqrt{4v_u^2 v_d^2+v_L^2 v^2}} \\ 
\end{array}\right],\\
& {\rm Diag}[m_{z_1}^2,m_{z_2}^2,m_{z_3}^2,m_{z_4}^2]
=
{\rm Diag}\left[0,0,0,-\lambda_0 \frac{v_u^2 v_L^2 + v_d^2 v'^2}{2 v_d v_u}\right],
\end{align}
where $v'\equiv \sqrt{v_L^2+ 4 v_u^2}$.
 Then they are respectively absorbed by neutral gauge bosons; $Z_{SM},Z_L,Z_R$.
In this model, the nature of THD sector is similar to that of type-II THDM. We thus omit detailed analysis of phenomenology in the scalar sector and focus on  gauge sector in our analysis below.

 \subsection{Neutral gauge boson sector}

{\it $Z_{SM}-Z_L-Z_R$ mixing}:
Since $H_{u,d}$ has nonzero $U(1)_R$ and $U(1)_L$ charges, there is mixing among  $Z_{SM},Z_L,Z_R$. 
The resulting mass matrix in basis of $(Z_{SM},Z_L,Z_R)$  is given by
\begin{align}
m_{Z_{SM},Z_L,Z_R}^2
&= 
\left[\begin{array}{ccc}
\frac{g^2 v^2}4 & -\frac{\ell g g_L (v_d^2-v_u^2)}{2} &  -\frac{r g g_R v^2}{2 }  \\ 
 -\frac{\ell g g_L (v_d^2-v_u^2)}{2} & \ell^2 g_L^2  (v^2+  v_L^2)  &   r\ell g_L g_R (v_d^2-v_u^2) \\
 -\frac{r g g_R v^2}{2}  & r\ell g_L g_R (v_d^2-v_u^2)  & r^2 g_R^2 (v^2+ 4 v_R^2)   \\ 
\end{array}\right],
\end{align}  
where $g^2\equiv g_1^2+g_2^2$, $m_{Z_{SM}}\equiv \frac{\sqrt{g_1^2+g_2^2}v}{2}\approx 91.18$ GeV, $g_1$, $g_2$, $g_L$, and $g_R$ are gauge coupling of $U(1)_Y$, $SU(2)_L$, $U(1)_L$, and $U(1)_R$, respectively.
Then its mass matrix is diagonalized by the three by three mixing matrix $V$ as $V m_{Z_{SM},Z_L,Z_R}^2 V^T
\equiv {\rm Diag}(m^2_{Z_1},m^2_{Z_{2}}, m^2_{Z_{3}})$, where $m^2_{Z_1}$ is identified as the measured neutral gauge boson. 
Here we find that $Z_L$ does not mix with the others when $v_u = v_d$.
Note that mixing between $Z_{SM}$ 
and the other neutral gauge bosons are strongly constrained.
We thus assume the mixings are negligibly small. This can be realized when $m_{Z_1} \ll m_{Z_{2,3}}$.
On the other hand mixing between $Z_L$ and $Z_R$ can be sizable if their masses are same order and we take the mixings
\begin{equation}
V_{11} \sim 1, \quad \{ V_{1a'}, V_{a'1} \} \ll 1, \quad V_{22} = V_{33} = \cos A, \quad V_{23} = -V_{32} = \sin A,
\label{eq:Vij}
\end{equation} 
where $a' = 2,3$ and we have introduced mixing angle $A$.
For the case of $m_{Z_1}(\simeq m_{Z_{SM}}) \ll m_{Z_{2,3}}$ the mixing angle $A$ is roughly given by
\begin{equation}
\sin 2 A \sim \frac{r \ell g_L g_R (v_d^2 - v_u^2)}{m^2_{Z_2} - m^2_{Z_3}}.
\label{eq:mixingA}
\end{equation}
Thus mixing is typically small unless $m_{Z_2} \sim m_{Z_3}$.
In addition, precise measurement of $Z$ boson mass would give strong constraint. 
Since the ambiguity of the $Z$ boson mass is around $0.0021$~\cite{Olive:2016xmw}, one has to require
\begin{align}
|m_{Z_{SM}}-m_{Z_1}|\lesssim 0.0021\ {\rm GeV}.
\label{eq:const_zm}
\end{align}       
Therefore stringent constraint in terms of mass parameters would be induced by mass eigenvalue of $m_{Z_1}$ and (\ref{eq:const_zm}).
Note that we can tune parameters in the model to satisfy the condition in contrast to the case with only $U(1)_R$~\cite{Nomura:2017tih} and the constraint on the new gauge boson masses is less stringent. In this paper, we thus just assume $m_{Z_1} \simeq m_{Z_{SM}}$ for simplicity.

\subsection{ Fermion sector}

{\it Fermion masses:} After spontaneous symmetry breaking, we find fermion masses as
$m_u=v_u(y_u)_{ab}/\sqrt2$, $m_d=v_d(y_d)_{ab}/\sqrt2$, and $m_\ell=v_d(y_\ell)_{ab}/\sqrt2$. 
The exotic fermions are also found to be $m_{Q'}(= m_{u'}= m_{d'})\equiv y_{Q'}v_L/\sqrt2$ and $m_{L'}(= m_{n'}= m_{e'})\equiv y_{L'}v_L/\sqrt2$, where we define $Q'\equiv [u',d']^t$ and $L'\equiv [n',e']^t$ respectively.
 
In the neutral sector, we have the six by six mass mass matrix in basis of $(\nu_L,\nu_R)$ as given by
\begin{align}
{\cal M}_N=
\left[\begin{array}{cc}
0 & m_D  \\ 
m_D^T  &  M_N  \\ 
\end{array}\right],
\end{align}
and ${\cal M}_N$ is diagonalized by $(D_{\nu_l},D_{\nu_H})\equiv O_N {\cal M}_N O_N^T$,
where $m_D\equiv y_D v/\sqrt2$, $M_N\equiv y_\nu v_R/\sqrt2$, and $O_N$ is six by six unitary matrix. 
Assuming $m_D<<M$, one finds the following mass eigenvalues and their mixing $O_N$~\cite{Huitu:2008gf}:
\begin{align}
 D_{\nu_l} & \equiv V_{MNS} m_\nu V^T_{MNS}
 \approx -2V_{MNS} m_D M^{-1} m_D^T V^T_{MNS} , \\
D_{\nu_H} & \approx M_N,\quad 
O_N\approx
\left[\begin{array}{cc}
V_{MNS} & 0  \\ 
0  &  1  \\ 
\end{array}\right]
\left[\begin{array}{cc}
-1 & \theta  \\ 
\theta^T  &  1  \\ 
\end{array}\right],
\end{align}
where $\theta\equiv m_D M^{-1}$, $V_{MNS}$ and $D_{\nu_l}$ are observable and fixed by the current neutrino oscillation data~\cite{Olive:2016xmw}.
One also finds the following relation between flavor- and mass-eigenstate:
\begin{align}
\nu_L\approx -V^T_{MNS} \nu_l + \theta \nu_H,\quad
\nu_R\approx -\theta^\dag V^\dag_{MNS} \nu_l + \nu_H.
\end{align}

{\it Gauge interactions for neutral fermions}:
Now that we formulate the masses and their mixings for the fermions,
one can write down the interactions from the kinetic term in Lagrangian under $SU(2)_L\times U(1)_Y\times U(1)_L\times U(1)_R$ gauge symmetry.
First of all, let us focus on neutral fermion sector.
Then one can write down their interactions in terms of mass eigenstate as
\begin{align} 
{\cal L}_{\nu}&\sim
\frac{g_2}{\sqrt2}
\left[W^-_\mu\bar\ell \gamma^\mu P_L(-V_{MNS}^T \nu_{l} +\theta \nu_{H} ) +{\rm h.c.}\right]\\
&+
\sum_{a=1}^3 Z^\mu_a
\left[\left(\frac{g}2 V^T_{1a}+\ell g_L V^T_{2a}\right) \left(\bar\nu_{l} V^*_{MNS} \gamma_\mu \theta P_L \nu_{H} + {\rm h.c.}\right)
+
\left(r g_R V^T_{3a}\right) \left(\bar\nu_{l} V_{MNS}\theta \gamma_\mu P_R \nu_{H} + {\rm h.c.}\right)
\right],\nn
\end{align}
where we do not consider the neutral component of $L'$, since it does not mix each other.

{\it The $Z_a$ interactions with charged fermions in the SM}:
The interactions associated with neutral gauge bosons and charged fermions in the SM are given by
\begin{align} 
- {\cal L}_{Z_a f \bar f}  = & \sum_{a = 1}^3 \bar f \gamma_\mu \biggl[ \left( -\frac{g_2}{c_W} (T^3 - s_W^2 Q_f ) V^T_{1a} + Q_{L_f} g_L V_{2a}^T \right) P_L  \nonumber \\
 & \qquad \qquad + \left( \frac{g_2}{c_W} (s_W^2 Q_f) V^T_{1a} + Q_{R_f} g_R V^T_{3a} \right) P_R \biggr] f Z_a^\mu \nonumber \\
 \simeq & \bar f \gamma_\mu \biggl[ \left( -\frac{g_2}{c_W} (T^3 - s_W^2 Q_f ) \right) P_L + \left( \frac{g_2}{c_W} (s_W^2 Q_f) \right) P_R \biggr] f Z_1^\mu  \\
 &+  \bar f \gamma_\mu \left[ Q_{L_f} g_L V^T_{22} P_L + Q_{R_f} g_R V^T_{32} P_R \right] f Z_2^\mu
 +  \bar f \gamma_\mu \left[ Q_{L_f} g_L V^T_{23} P_L + Q_{R_f} g_R V^T_{33} P_R \right] f Z_3^\mu,\nn
 \label{eq:intZpff}
 \end{align}
where $f$ indicates charged leptons and quarks in the SM, $Q_f$ is the electric charge of $f$, $Q_{L_f(R_f)}$ is the $U(1)_{L(R)}$ charge of $f$, $s_W(c_W) = \sin \theta_W (\cos \theta_W)$ with Weinberg angle $\theta_W$ and we have applied the approximation $V_{11} \sim 1 (V_{12,13,21,31} \ll 1)$ in the second equality.

{\it The $Z_a$ interactions with exotic fermions }: 
Furthermore, the interactions associated with the exotic fermions are respectively given by
\begin{align} 
{\cal L}_{L'}& = \sum_{a=1}^3   \left[
\bar n'\gamma_\mu \left(  \frac{g_2}{c_W} V^T_{1a} +\ell g_L V^T_{2a} P_R \right) n' 
+ \bar \ell'\gamma_\mu \left(  - \frac{g_2}{c_W} \left(\frac{1}{2} - s_W^2 \right) V^T_{1a} +\ell g_L V^T_{2a} P_R \right) \ell' \right]  Z_{a}^\mu, \nn\\
{\cal L}_{Q'}& = \sum_{a=1}^3 \biggl[
\bar u'\gamma_\mu 
\left( \frac{g_2}{c_W} \left(\frac12 - \frac23 s_W^2 \right)  V^T_{1a} +\ell g_L V^T_{2a} P_R\right) u' \nonumber \\
& \qquad  \qquad - \bar d'\gamma_\mu 
\left(  \frac{g_2}{c_W} \left(\frac12 + \frac13 s_W^2 \right) V^T_{1a} +\ell g_L V^T_{2a} P_R \right) d'  \biggr] Z_{a}^\mu.
\end{align}
In addition, electromagnetic interactions are the same structure as the SM one; ${\cal L}_{\gamma f' \bar f'} = -e\bar\ell'\gamma^\mu\ell'A_\mu+\frac{2e}{3}\bar u'\gamma^\mu u'A_\mu - \frac{e}{3}\bar d'\gamma^\mu d'A_\mu$.

\section{$Z'$ bosons at colliders}

In this section, we discuss collider physics of $Z'$ bosons focusing on interactions with the SM fermions.

\subsection{$Z'$ boson productions in proton-proton collider}

Our exotic neutral gauge bosons $Z_{2,3}$ can be produced at the LHC through the $\bar q  q \to Z_{2,3}$ process, since they couple to the SM quarks.
The gauge interactions among $Z_{2,3}$ and SM quarks are given by
\begin{align}
- {\cal L} \supset & \bar u^a \gamma_\mu \left[ \ell g_L \cos A P_L + r g_R \sin A P_R \right] u^a Z_2^\mu
 -  \bar u^a \gamma_\mu \left[  \ell g_L \sin A P_L + r g_R \cos A P_R \right] u^a Z_3^\mu \nn \\
 & + \bar d^a \gamma_\mu \left[ \ell g_L \cos A P_L - r g_R \sin A P_R \right] d^a Z_2^\mu
 -  \bar d^a \gamma_\mu \left[  \ell g_L \sin A P_L - r g_R \cos A P_R \right] d^a Z_3^\mu,
\end{align}
where we have used Eq.~(\ref{eq:Vij}) for mixing matrix $V$. 
We note that interaction for SM charged leptons is the same form as that of $d$-quarks.  
The $Z_{2,3}$ decay into quarks and leptons with BRs as BR$(Z_{2,3} \to \bar q q) \simeq 3 {\rm BR}(Z_{2,3} \to \ell^+ \ell^-)$ due to universal couplings to quark and lepton sectors~\footnote{Here we assume exotic fermions are heavier than $m_{Z_{2,3}}/2$ to forbid $Z_{2,3}$ from decaying into them for simplicity.}.
The strongest constraint on the $Z_{2,3}$ masses and couplings is derived by searching for $pp \to Z_{2,3} \to \ell^+ \ell^- (\ell^\pm = e^\pm, \mu^\pm)$ processes.
Estimating the cross section with {\it CalcHEP}~\cite{Belyaev:2012qa} implementing relevant interactions and using the CTEQ6 parton distribution functions (PDFs)~\cite{Nadolsky:2008zw}, we obtain the cross section of the processes as $\sim 0.06$ fb with $m_{Z_{2(3)}} = 4$ TeV, $g_{L(R)} = 0.1$ and $\sin A \ll 1$.
This cross section is marginal to the current constraint and the masses of $Z_{2,3}$ should be TeV scale or larger when $g_{L,R} \geq \mathcal{O}(0.1)$.
Note that chiral structure of the gauge interactions could be investigated by measuring angular distributions of lepton plus jets final states coming from $t \bar t$ pair via $Z_{2,3}$~\cite{Cerrito:2016qig}.

Here we also note that exotic fermions $Q'$, $L'$ as well as $\nu_R$ can be produced through interaction with $Z_{2,3}$. In this paper we just assume these particles are sufficiently  heavy and further discussion is left in future work since we focus on signal of $Z_{2,3}$ bosons.

\subsection{Test of $Z_{2,3}$ interaction at $e^+ e^-$ collider}

At the $e^+ e^-$ collider, on-shell  $Z_{2,3}$ bosons will not be directly produced if the mass of $Z_{2,3}$ are few TeV scale or heavier.
However we can test the interactions among $Z_{2,3}$ and charged leptons by measuring deviation from the SM prediction in the scattering processes $e^+ e^- \to \ell^+ \ell^-$.
These scattering processes can be considered using effective operator approach for $s \ll m_{Z_{2,3}}^2$.
In our model, effective operators can be written by
\begin{align}
&{\cal L}_{eff} 
= \sum_{\ell' = e, \mu, \tau} \sum_{X = L, R} \sum_{X' = L, R}  \frac{4 \pi}{1+ \delta_{e \ell'}} 
\left[ \frac{1}{(\Lambda^{\ell'}_{XX'})^2} (\bar e \gamma^\mu P_X e)(\bar \ell' \gamma_\mu P_{X'} \ell') \right],  \nonumber \\
& (\Lambda_{LL}^{\ell'} )^{-1}\equiv g_L \ell \sqrt{ \frac{1}{2 \pi (1+\delta_{e \ell'})} \left( \frac{\cos^2 A}{m^2_{Z_2}} + \frac{\sin^2 A}{m^2_{Z_3}} \right)}, \nonumber \\
& (\Lambda_{RR}^{\ell'})^{-1} \equiv g_R r \sqrt{ \frac{1}{2 \pi (1+\delta_{e \ell'})} \left( \frac{\sin^2 A}{m^2_{Z_2}} + \frac{\cos^2 A}{m^2_{Z_3}} \right) }, \nonumber \\
& (\Lambda_{LR}^{\ell'})^{-1} = (\Lambda_{RL}^{\ell'})^{-1} \equiv  \sqrt{ \frac{1}{2 \pi (1+\delta_{e \ell'})} \left| r \ell g_L g_R \cos A \sin A \left( \frac1{m_{Z_2}^2} - \frac1{m_{Z_3}^2} \right) \right|} ,
\end{align}
where $\delta_{e\ell'}$ is the Kronecker delta, and we have applied interactions in Eq.~(\ref{eq:intZpff}) and used Eq.~(\ref{eq:Vij}) for mixing matrix $V$.
Here we note that $(\Lambda_{LR}^{\ell'})^{-1}$ is suppressed by $\sqrt{r \ell g_L g_R (v_d^2 - v_u^2)}/m_{Z_{2,3}}$ compared to $\Lambda_{LL(RR)}^{\ell'}$ due to the mixing angle $A$ given by Eq.~(\ref{eq:mixingA}); either $|\sin A|$ or $|1/m_{Z_2}^2 - 1/m_{Z_3}^2|$ factor is small in the square root.
Thus we ignore $\Lambda_{LR(RL)}^{\ell'}$ in the following analysis as an approximation. 
Furthermore we take $\ell=r=1$ to reduce the number of free parameter.

We can test dependence of scattering processes on left-handed and right-handed types of interactions by an analysis with polarized initial state at the ILC.
To apply the method discussed in ref.~\cite{Nomura:2017abh}, we consider the processes
\begin{align}
e^-(k_1,\sigma_1)  e^+(k_2,\sigma_2) \to \ell^-(k_3,\sigma_3)  \ell^+(k_4,\sigma_4),
\end{align}
where $k_i$ indicates 4-momentum of each particle and we explicitly show the helicities of initial- and final-state leptons $\sigma_{i} = \pm$.
Combining the SM and $Z_{2,3}$ contributions, helicity amplitudes ${\cal M}_{\sigma_i} = {\cal M}(\sigma_1  \sigma_2 \sigma_3 \sigma_4)$ 
for the process of $e^-(\sigma_1)  e^+(\sigma_2) \to e^-(\sigma_3)  e^+(\sigma_4)$ are given by
\begin{align}
  & {\cal M}(+-+-) = -e^2\left(1+\cos\theta\right)
 \left[ 1 + \frac{s}{t} + c_R^2\left(\frac{s}{s_Z}+\frac{s}{t_Z}\right)
 + \frac{2 s}{\alpha (\Lambda_{RR}^e)^2}\right], \\ 
 & {\cal M}(-+-+) = -e^2\left(1+\cos\theta\right)
 \left[ 1 + \frac{s}{t} + c_L^2\left(\frac{s}{s_Z}+\frac{s}{t_Z}\right)
 + \frac{2 s}{\alpha (\Lambda_{LL}^e)^2}\right], \\ 
 & {\cal M}(+--+) = {\cal M}(-++-) =
 e^2\left(1-\cos\theta\right)\left[1+c_Rc_L\frac{s}{s_Z}\right], \\
 & {\cal M}(++++) = {\cal M}(----) =
 2e^2\frac{s}{t}\left[1+c_Rc_L\frac{t}{t_Z}\right],
\end{align}
where $t=(k_1-k_3)^2=(k_2-k_4)^2=-s(1-\cos\theta)/2$, $s=(k_1+k_2)^2=(k_3+k_4)^2$, 
$s_Z=s-m_Z^2+im_Z\Gamma_Z$, $t_Z=t-m_Z^2+im_Z\Gamma_Z$, 
$e^2=4\pi\alpha$ with $\alpha$ being the QED coupling constant,
$c_R=\tan\theta_W$, $c_L=-\cot2\theta_W$,
and $\cos\theta$ is the scattering polar angle. 
The helicity amplitudes for $e^+ e^- \to \mu^+ \mu^-(\tau^+ \tau^-)$ are obtained by removing terms with $1/t$ and $1/t_Z$ and replacing $\Lambda_{L,R}^e$ by $\sqrt{2} \Lambda_{L,R}^{\mu(\tau)}$. Note also that, in the following analysis, we omit the case of $\tau^+ \tau^-$ final state since it is less sensitive compared to the others.

Applying the amplitudes, the differential cross-section for purely-polarized initial-state $\sigma_{1,2} = \pm1$, is obtained as 
\begin{align}
 \frac{d\sigma_{\sigma_1\sigma_2}}{d\cos\theta} = \frac{1}{32\pi s}
 \sum_{\sigma_3,\sigma_4} \left|{\cal M}_{\{\sigma_i\}}\right|^2.
\end{align}
Then we define partially-polarized differential cross section such that
\begin{align}
\frac{d \sigma (P_{e^-}, P_{e^+})}{d \cos \theta} = \sum_{\sigma_{e^-}, \sigma_{e^+} = \pm} \frac{1+ \sigma_{e^-} P_{e^-}}{2} \frac{1 +\sigma_{e^+} P_{e^-}}{2} \frac{d \sigma_{\sigma_{e^-} \sigma_{e^+}}}{d \cos \theta},
\end{align}
where $P_{e^-(e^+)}$ is the degree of polarization for the electron(positron) beam and the helicity of final states is summed up. 
Polarized cross sections $\sigma_{L,R}$ are also defined by the following two cases as realistic values at the ILC~\cite{Baer:2013cma}:
\begin{equation}
\frac{d \sigma_{R}}{d \cos \theta} = \frac{d \sigma (0.8,-0.3)}{d \cos \theta}, \quad \frac{d \sigma_{L}}{d \cos \theta} = \frac{d \sigma (-0.8,0.3)}{d \cos \theta}.
\end{equation}
We apply the polarized cross sections to study the sensitivity to $Z_{2,3}$ bosons in $e^+ e^- \to \ell^+ \ell^-$ scattering 
via the measurement of a forward-backward asymmetry at the ILC, which is given by 
\begin{align}
& A_{FB} = \frac{N_F - N_B}{N_F + N_B}, \nonumber \\
& N_{F(B)} = \epsilon L \int_{0(-c_{\rm max})}^{c_{\rm max}(0)} d \cos \theta \frac{d \sigma}{d \cos \theta},
\end{align}
where $L$ is an integrated luminosity, a kinematical cut $c_{max}$ is chosen to maximize the sensitivity, and $\epsilon$ is an efficiency depending on the final states.
In our analysis we assume $\epsilon = 1$ for electron and muon final states, and $c_{\rm max} = 0.5(0.95)$ is taken for electron(muon) final state~\cite{Tran:2015nxa}.
Then the forward-backward asymmetry is estimated for cases with only the SM gauge boson contributions, and with both SM and $Z_{2,3}$ boson contributions,
in order to investigate the sensitivity to $Z_{2,3}$.
Therefore the former case gives $N_{F(B)}^{SM}$ and $A_{FB}^{SM}$ while the latter case $N_{F(B)}^{SM+Z_2 + Z_3}$ and $A_{FB}^{SM+Z_2 + Z_3}$.    
The sensitivity to $Z_{2,3}$ interaction is thus estimated by 
\begin{equation}
\Delta A_{FB}(\sigma_{L,R}) = |A_{FB}^{SM+Z_{2}+Z_{3}} (\sigma_{L,R})- A_{FB}^{SM}(\sigma_{L,R})|.
\label{eq:delAFB}
\end{equation}
We compare this quantity with a statistical error of the asymmetry, assuming only SM contribution
\begin{equation}
\delta_{A_{FB}}^{SM} = \sqrt{\frac{1-(A_{FB}^{SM})^2}{N_F^{SM}+N_B^{SM}}},
\label{eq:SMerror}
\end{equation}
where both $\sigma_L$ and $\sigma_R$ cases are considered separately.

\begin{figure}[t]
\begin{center}
\includegraphics[width=70mm]{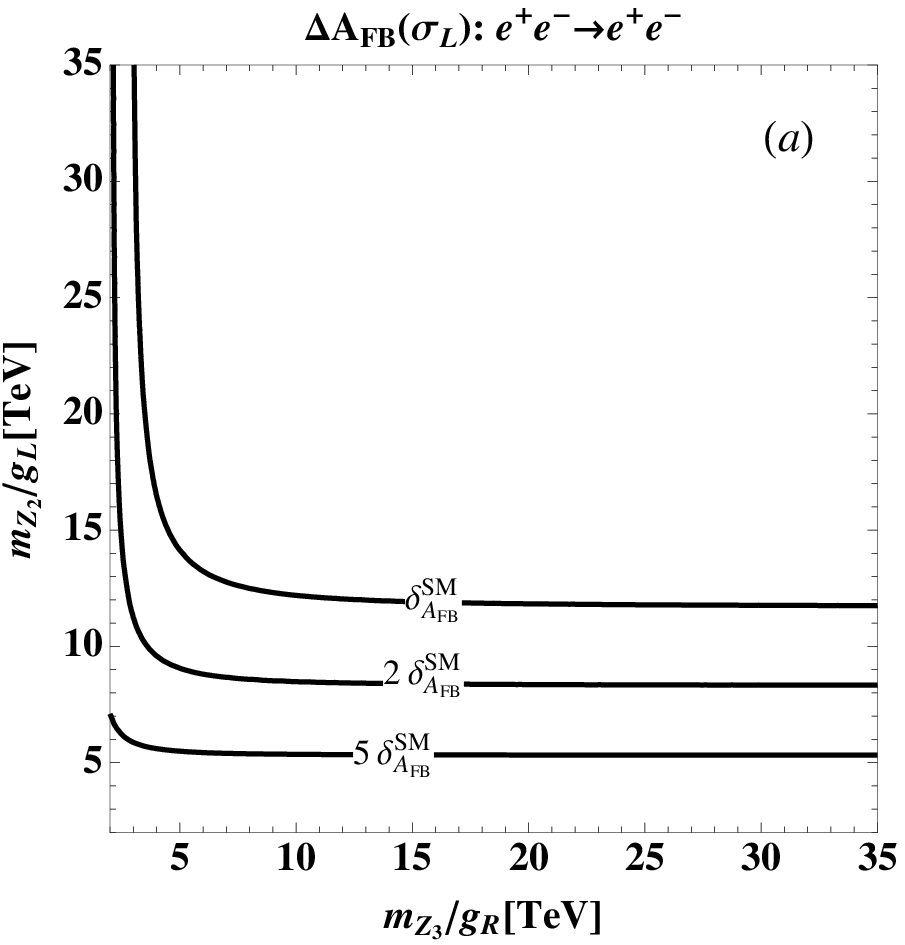} \quad
\includegraphics[width=70mm]{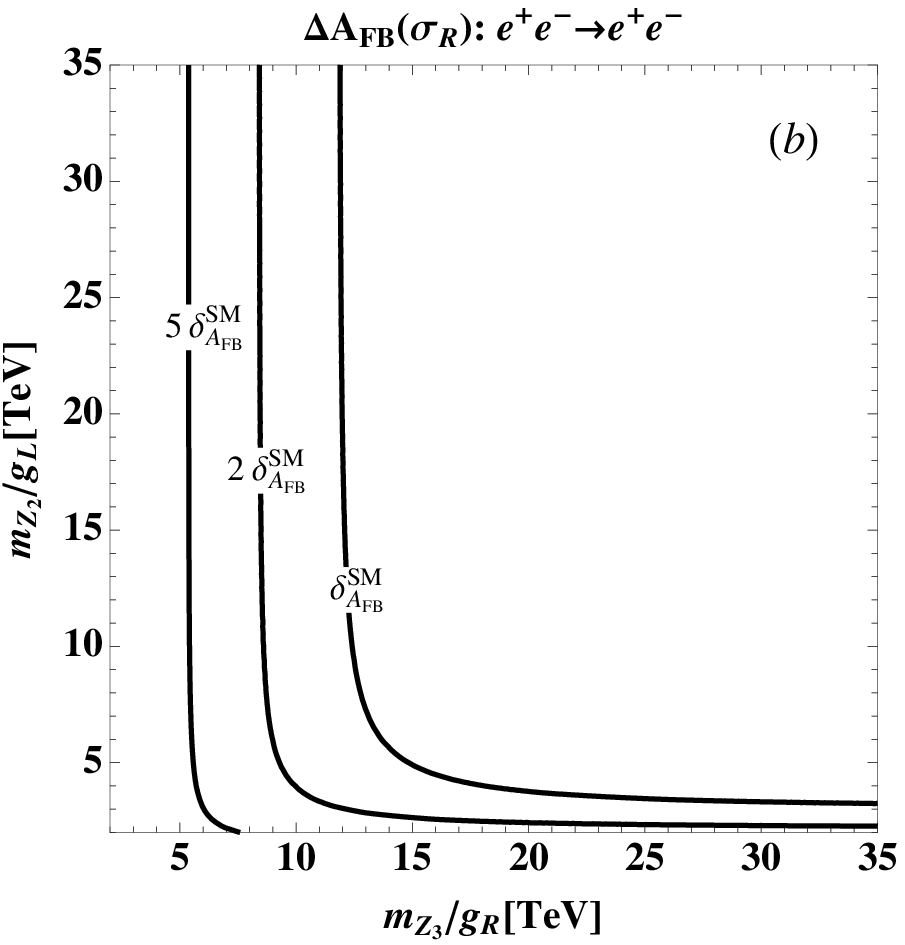}
\caption{The contours of $\Delta A_{FB}$ defined as Eq.~(\ref{eq:delAFB}) on the $m_{Z_2}/g_L$-$m_{Z_3}/g_R$ plane for $e^+e^- \to e^+ e^-$ process where we have assumed $\sin A \ll 1$ and applied the polarized cross section $\sigma_R$ and $\sigma_L$ for (a) and (b). The statistical error in the SM, $\delta_{A_{\rm FB}}^{\rm SM}$ given by Eq.~(\ref{eq:SMerror}), is estimated to be $3.67 \times 10^{-3}$ and $3.58 \times 10^{-3}$ for $\sigma_R$ and $\sigma_L$ respectively.} 
  \label{fig:AFBe}
\end{center}\end{figure}

\begin{figure}[t]
\begin{center}
\includegraphics[width=70mm]{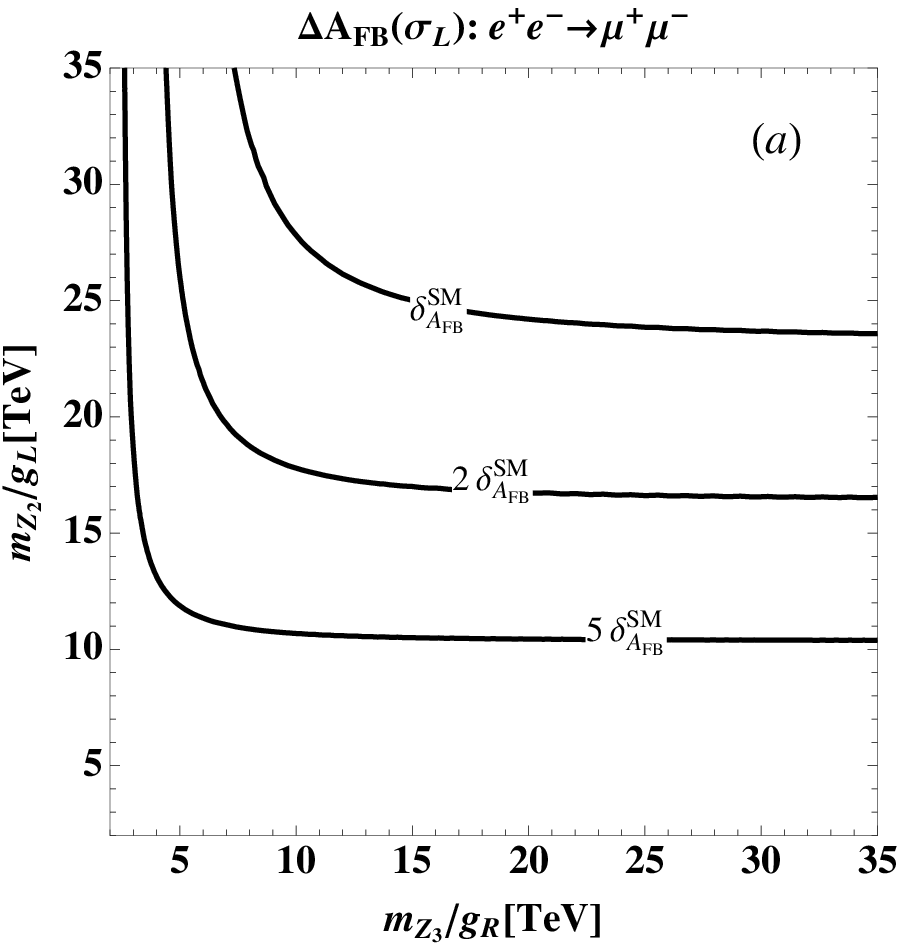} \quad
\includegraphics[width=70mm]{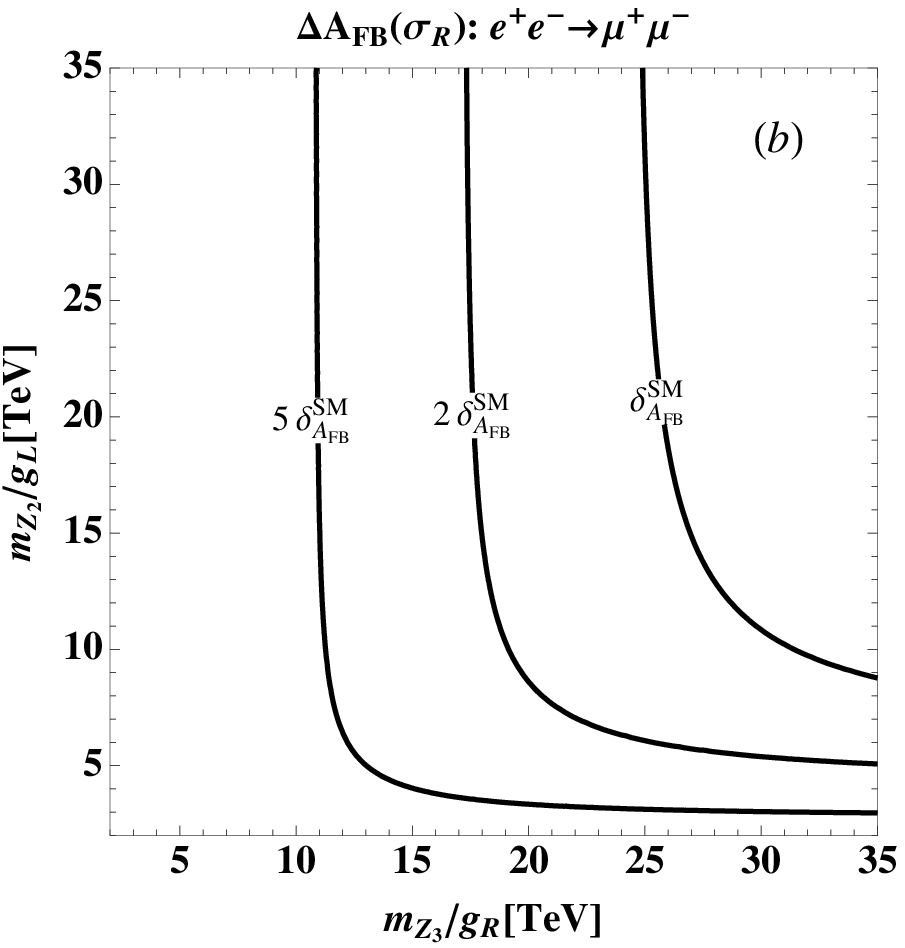}
\caption{The contours of $\Delta A_{FB}$ for the process $e^+e^- \to \mu^+ \mu^-$ where the other settings are the same as Fig.~\ref{fig:AFBe}. The statistical error in the SM, $\delta_{A_{\rm FB}}^{\rm SM}$ given by Eq.~(\ref{eq:SMerror}), is estimated to be $6.53 \times 10^{-3}$ and $5.73 \times 10^{-3}$ for $\sigma_R$ and $\sigma_L$ respectively.} 
  \label{fig:AFBm}
\end{center}\end{figure}

Fig.~\ref{fig:AFBe}-(a) and -(b) show the contours of $\Delta A_{FB}(\sigma_{L})$ and $\Delta A_{FB}(\sigma_{R})$ for the $e^+ e^- \to e^+ e^-$ process applying $\sin A \ll 1$, $\sqrt{s} = 250$ GeV and integrated luminosity of 1000 fb$^{-1}$. The contours show the values $5 \delta_{A_{FB}}^{SM}$, $2 \delta_{A_{FB}}^{SM}$ and $\delta_{A_{FB}}^{SM} \simeq 3.67(3.58) \times 10^{-3}$ for $\sigma_R (\sigma_L)$. 
From the contour plots we clearly see that $m_{Z_2}/g_L$ and $m_{Z_3}/g_R$ are respectively sensitive to the forward-backward asymmetry obtained from $\sigma_L$ and $\sigma_R$. 
We thus use the analysis with polarized beam to test the two types of gauge couplings by comparing the results from $\sigma_L$ and $\sigma_R$.
Also the effective coupling up to scale of $m_{Z_{2(3)}}/g_{L(R)} \sim 10$ TeV can be tested with $2 \sigma$ level by data from sufficient integrated luminosity.
Remarkably if the gauge couplings are not so small we can even test $Z_{2,3}$ masses heavier than the mass which can be directly produced at the LHC.
In addition, we show the case of $e^+ e^- \to \mu^+ \mu^-$ process in Fig.~\ref{fig:AFBm} where $\delta_{A_{FB}}^{SM} \simeq 6.23(5.73) \times 10^{-3}$ for $\sigma_R(\sigma_L)$ is estimated in this case. These plots indicate the sensitivity that is stronger than the case of $e^+ e^- \to e^+ e^-$ scattering and the scale of $m_{Z_{2(3)}}/g_{L(R)} \sim 20$ TeV can be tested with $2 \sigma$ level.
Here we also consider the case of large mixing $\sin A \sim 1/\sqrt{2}$ with $m_{Z_2} = m_{Z_3}$. In such a case, we obtain the same figure as Figs.~\ref{fig:AFBe} and \ref{fig:AFBm} by taking $m_{Z_2} = m_{Z_3} \equiv M$. In any cases, gauge couplings $g_L$ and $g_R$ are respectively sensitive to analysis with $\sigma_L$ and $\sigma_R$.  

\section{Summary and Conclusions}
 We have proposed a model with the left-handed and right-handed continuous Abelian gauge symmetry $U(1)_L \times U(1)_R$ introducing several exotic field contents as a minimal construction of such gauge theory.
Then we have introduced exotic quarks and leptons in order to cancel the new gauge anomalies, two Higgs doublet fields to induce nonzero SM fermion masses and two SM singlet scalar fields with new $U(1)$ charges to break the additional $U(1)$ gauge symmetries and to provide masses of exotic fermions.
Then we have formulated each of fermion sector, Higgs boson sector, vector gauged boson sector, as well as neutrino sector.
We have found the Yukawa interaction among two Higgs doublets and the SM fermions is that of type-II two Higgs doublet model.
Also active neutrino masses can be obtained after spontaneous gauge symmetry breaking as the same way as type-I seesaw mechanism.

As a direct result of two gauge symmetries, their breaking scales can be within several TeV which is lower than the case of only $U(1)_R$ symmetry; this is due to additional parameters and degrees of freedom in the gauge boson sector. 
In addition we have discussed the possibility of testing the new gauge interactions associated with new gauge bosons, $Z_{2,3}$, at current and future  collider such as the LHC and the ILC and of distinguishing differences between our model and the others. The exotic neutral gauge bosons can be directly produced at the LHC, since they couple to the SM quarks. Then the strongest constraint is obtained from the mode in which produced $Z_{2,3}$ decays into SM charged leptons, and $Z_{2,3}$ should be heavier than $\sim 4$ TeV when the new gauge coupling is more than $\mathcal{O}(0.1)$.
In particular, we have shown that the chiral structure of gauge interactions can be investigated by the analysis of forward-backward asymmetry based on polarized electron(positron) beam at the ILC.  It is found that $\sim 10-20$ TeV scale of $m_{Z_{2,3}}/g_{L,R}$ can be tested with the ILC data from $\sqrt{s} = 250$ GeV and integrated luminosity of 1000 fb$^{-1}$. Furthermore since sensitivity to left- and right-handed types of gauge interactions depends on type of polarized beam, we can distinguish which types of interaction is stronger than the others.


\section*{Acknowledgments}
H. O. is sincerely grateful for KIAS and all the members.


\begin{thebibliography}{99}

\bibitem{Geng:1993ja} 
  C.~Q.~Geng, K.~Whisnant and B.~L.~Young,
  hep-ph/9302273.

\bibitem{Nomura:2016emz} 
  T.~Nomura and H.~Okada,
  Phys.\ Lett.\ B {\bf 761}, 190 (2016)
  [arXiv:1606.09055 [hep-ph]].

\bibitem{Nomura:2016pgg} 
  T.~Nomura and H.~Okada,
  Phys.\ Rev.\ D {\bf 94}, no. 9, 093006 (2016)
  [arXiv:1609.01504 [hep-ph]].
  
\bibitem{Nomura:2017ezy} 
  T.~Nomura and H.~Okada,
  Phys.\ Rev.\ D {\bf 96}, no. 1, 015016 (2017)
  [arXiv:1704.03382 [hep-ph]].

\bibitem{Nomura:2017tih} 
  T.~Nomura and H.~Okada,
  arXiv:1707.00929 [hep-ph].
  
\bibitem{Chao:2017rwv} 
  W.~Chao,
  arXiv:1707.07858 [hep-ph].

\bibitem{Mohapatra:1980qe} 
  R.~N.~Mohapatra and R.~E.~Marshak,
  Phys.\ Rev.\ Lett.\  {\bf 44}, 1316 (1980)
  Erratum: [Phys.\ Rev.\ Lett.\  {\bf 44}, 1643 (1980)].
  
    \bibitem{ilc} 
  S. Riemann, LC-TH-2001-007.



\bibitem{Nomura:2017tzj} 
  T.~Nomura and H.~Okada,
  arXiv:1706.05268 [hep-ph].
  
    
   \bibitem{Gunion:1989we} 
  J.~F.~Gunion, H.~E.~Haber, G.~L.~Kane and S.~Dawson,
  ``The Higgs Hunter's Guide,''
  Front.\ Phys.\  {\bf 80}, 1 (2000).
  

  
  
\bibitem{Huitu:2008gf} 
  K.~Huitu, S.~Khalil, H.~Okada and S.~K.~Rai,
  Phys.\ Rev.\ Lett.\  {\bf 101}, 181802 (2008)
  [arXiv:0803.2799 [hep-ph]].
  
\bibitem{Olive:2016xmw} 
  C.~Patrignani {\it et al.} [Particle Data Group],
  Chin.\ Phys.\ C {\bf 40}, no. 10, 100001 (2016).
  
    \bibitem{Belyaev:2012qa} 
  A.~Belyaev, N.~D.~Christensen and A.~Pukhov,
  Comput.\ Phys.\ Commun.\  {\bf 184}, 1729 (2013)
  [arXiv:1207.6082 [hep-ph]].
  
\bibitem{Nadolsky:2008zw} 
  P.~M.~Nadolsky, H.~L.~Lai, Q.~H.~Cao, J.~Huston, J.~Pumplin, D.~Stump, W.~K.~Tung and C.-P.~Yuan,
  Phys.\ Rev.\ D {\bf 78}, 013004 (2008)
  [arXiv:0802.0007 [hep-ph]].

\bibitem{Cerrito:2016qig} 
  L.~Cerrito, D.~Millar, S.~Moretti and F.~Spano,
  arXiv:1609.05540 [hep-ph].
  
\bibitem{Nomura:2017abh} 
  T.~Nomura, H.~Okada and H.~Yokoya,
  arXiv:1702.03396 [hep-ph].

\bibitem{Baer:2013cma} 
  H.~Baer {\it et al.},
  arXiv:1306.6352 [hep-ph].
  
\bibitem{Barklow:2015tja} 
  T.~Barklow, J.~Brau, K.~Fujii, J.~Gao, J.~List, N.~Walker and K.~Yokoya,
  arXiv:1506.07830 [hep-ex].
  
\bibitem{Tran:2015nxa} 
  T.~H.~Tran, V.~Balagura, V.~Boudry, J.~C.~Brient and H.~Videau,
  Eur.\ Phys.\ J.\ C {\bf 76}, no. 8, 468 (2016)
  [arXiv:1510.05224 [physics.ins-det]].
  
  
  
\end{thebibliography}
\end{document}